\documentclass[twocolumn,floatfix,aps,prl,showpacs]{revtex4}
\usepackage{graphicx}

\begin{document}

\title{Relaxation pathway confinement in glassy dynamics}

\author{J. A. Rodriguez Fris, M. A. Frechero and G. A. Appignanesi}

\affiliation{Secci\'on Fisicoqu\'{\i}mica, INQUISUR-UNS-CONICET,  Universidad Nacional del Sur, Av. Alem 1253, 8000 Bah\'{\i}a Blanca, Argentina.\\}

\date{\today}

\begin{abstract}
We compute for an archetypical glass-forming system the excess of particle mobility distributions over the corresponding distribution of dynamic propensity, a quantity that measures the tendency of the particles to be mobile and reflects the local structural constraints. This enables us to demonstrate that, on supercooling, the dynamical trajectory in search for a relaxation event must deal with an increasing confinement of relaxation pathways. This ``entropic funnel'' of relaxation pathways built upon a restricted set of mobile particles is also made evident from the decay and further collapse of the associated Shannon entropy.
\end{abstract}

\pacs{61.20.Ja; 61.20c.Lc; 64.70.Pf}

\maketitle

\section{Introduction}
The nature of the glass transition, that is, the dramatic dynamic slowing down that a glass-forming system experiences when supercooled below its melting point, continues to be one of the major open questions in condensed-matter physics \cite{angell_91,ediger_00,sastry_98,harrowellnatphys,chandlerscience,biroli_06,topicalreview}. A main ingredient for glassy dynamics is the presence of dynamical heterogeneities \cite{angell_91,ediger_00,donati_98,appignanesi_05}. According to this description, the relaxation slows down by a particle-caging effect whose lifetime enormously increases as temperature is lowered within the supercooled regime towards the mode-coupling temperature, $T_{\rm c}$, while the distribution of particle mobility becomes inhomogeneous or non-gaussian \cite{donati_98}. Additionally, the structural relaxation (or $\alpha$ relaxation) has been shown to be accomplished by means of sporadic bursts in mobility comprising the collective motion of a significant number of particles involved in a relatively compact cluster, or d-cluster \cite{appignanesi_05,topicalreview,vallee_07,weeks}. %These events, which represent natural candidates for the cooperatively relaxing regions proposed long ago by Adam and Gibbs, have been identified by computer simulations in different glass-forming systems \cite{appignanesi_05,waterPRE} and confirmed by confocal microscopy experiments \cite{vallee_07,weeks}. Additionally, they have been related to theoretical frameworks like inhomogeneous mode-coupling theory \cite{biroli_06} and dynamic facilitation \cite{garrahanjpa,chandlerscience}. 
In turn, the configuration-space counterpart of this real-space description is provided by the landscape paradigm \cite{sastry_98,appignanesi_05}: At low temperatures such that equilibration within local minima is fast compared to transitions between them, the dynamics of the system can be described as an exploration of its Potential Energy Landscape, PEL. The PEL is made up by an extremely large collection of local minima called inherent structrures, ISs, which in turn are arranged in superstructures called Metabasins, MBs. A MB is a group of closely related or similar ISs which is separated from other MBs by a long range particle rearrangement \cite{sastry_98,appignanesi_05,doliwaPRL_03}. Contrasting with the situation at higher temperatures when the system performs a non-glassy free-diffusion given that thermal energy is high enough to wash up the barriers of the PEL, the regime that sets in upon supercooling has been termed ``landscape influenced'' \cite{sastry_98} since the trajectory is subject to confinement within the local MB, with MB transitions triggered by a d-cluster event \cite{appignanesi_05,JCP,vallee_07,weeks}. Approaching $T_{\rm c}$, the dynamical trajectory gets increasingly MB-confined (MB-residence times grow significantly) and thus, the exploration of the PEL becomes ``landscape dominated'' \cite{sastry_98}. Additionally, a third main brush-stroke for the description of glassy dynamics came with the introduction of the concept of dynamic propensity. This quantity represents the tendency of the particles to be mobile, reflecting the constraints imposed by the local structure on the dynamics \cite{widmer-cooper_04,harrowellnatphys}. The distribution of particle propensities has been shown to be sharp at high temperature but to get progressively broader with supercooling \cite{widmer-cooper_04,harrowellnatphys,ludo}, when it becomes spatially heterogeneous and displays high-propensity particles arranged in relatively compact clusters (which have also been related to the local soft modes of the sample) \cite{widmer-cooper_04,PRL_2,harrowellnatphys}. It has also been shown that the propensity pattern decorrelates by means of d-cluster events (i.e., by MB transitions) and, thus, the influence of the structure on dynamics extends only at the local MB level while the long-time (difussive) dynamics of the system can be described as a random walk on MBs \cite{PRL_2,doliwaPRL_03,JCP}.
Under the above-expounded scenario, the aim of the present work is to learn how such brush-strokes add up in a big picture, so as to contribute to the understanding of the physical underpinnings of glassy relaxation. In other words, we wish to link dynamical slowing down, landscape influence and structural constraints as glassiness sets by decreasing temperature within the supercooled regime and the dynamical trajectory in search for a relaxation event (and, thus, for a MB transition) is confronted with a progressively more heterogeneous distribution of structural constraints, as reflected by the dynamic propensity. 

\section{Methodology}
We performed a series of molecular dynamics (MD) simulations within the NVE ensemble for a paradigm model of fragile glass former: the binary Lennard-Jones system consisting of a three-dimensional mixture of $80 \%$ A and $20 \%$ B particles, the size of the A particles ($\sigma_{AA}$) being $10 \%$ larger than the B ones (see \cite{appignanesi_05} for details; we note that all the quantities we calculate below involve only the A particles, on average these particles are slower than the B particles). We carried out simulation runs after equilibration for a range of temperatures from $T=3.0$ to $T=0.446$ and a density of $1.2$ \cite{appignanesi_05} (the Mode Coupling temperature for this system has been estimated to be $T_{\rm c}=0.435$). The systems were always equilibrated for timescales much larger than the correpsonding $\alpha$-relaxation time, $\tau_{\alpha}$ (for the lowest temperatures we equilibrated for timescales almost two orders of magnitude larger than $\tau_{\alpha}$). We also carefully tested equilibration by monitoring thermodynamic and dynamical quantities in order to discard drifts during the runs. For example, the value of $\tau_{\alpha}$ measured at different times during the runs after equilibration does not present any significant change (for systems that are not properly equilibrated $\tau_{\alpha}$ is expected to increase with time towards its equilibrium value since the system relaxes faster when not equilibrated). We employed $N=1000$ (but we note that systems of size $N=8000$ gave similar results). % The dynamics is most heterogeneous at a time $t^*$ defined by the maximum in the non-gaussian parameter, $\alpha_2(t) = \frac{3 \langle r^4(t)\rangle}{5 \langle r^2(t)\rangle ^2} - 1$, which measures the deviation of the self part of the van Hove function, the probability at a given time of finding a particle at distance $r$ from its initial position, from a brownian behavior (see \cite{donati_98,appignanesi_05} for details). This quantity is located at the end of the $\beta$ and the beginning of the $\alpha$ relaxation (the crossover from the caging to the diffusive regime, when the system leaves the plateau that appears, due to the caging effect, in the mean squared displacement function) and constitutes the characteristic time for dynamical heterogeneities. Additionally, $t^*$ depends strongly on temperature growing quickly as we move towards $T_{\rm c}$ \cite{donati_98,appignanesi_05} and has been shown to provide an estimate of the MB residence time \cite{appignanesi_05}.
%Once a system was equilibrated, we performed a single MD simulation for a total time of $1600 \cdot t^*$. $t^*$ at $T=0.446$ was estimated from local simulations to be 6000. The remaining $t^*(T)$ were estimated from reference \cite{kob_95}, that is, $t^*$ for $T=3$, $T=2$, $T=1$, $T=0.80$, $T=0.60$, $T=0.50$, $T=0.466$ is 3, 5, 15, 30, 90, 400 and 2500 respectively. @@@T2 y T3 las estimamos nosotros tmb@@@
%For our NVE simulations we need to keep the total energy E of the system constant. Since this is difficult for long runs, we affected the velocities of the particles every $10^6$ simulation steps by a factor ($\approx 1$) so as to get a kinetic energy that, when added to the potential energy, gives the initial total energy of the system. This procedure avoids a drift in E and keeps the averaged $T$ to its desired value.
%For each $T$, taking the corresponding long simulation of $1600 \cdot t^*$, we stored 160 configurations separated one another by $10\cdot t^*$ and from each one performed an isoconfigurational ensembles (IC).

To calculate propensities we employed the iso-configurational ensemble (IC) method introduced in \cite{widmer-cooper_04} by performing a series of equal length MD runs from the same initial configuration, that is, always the same structure (the same particle positions) but each one with different initial particle momenta chosen at random from the appropriate Maxwell-Boltzmann velocity distribution. We then calculated the dynamic propensity as \cite{widmer-cooper_04} $P_i = \langle \Delta {\bf r}_i^2(t)\rangle_{\rm IC}$, (where $\langle \cdots \rangle_{\rm IC}$ indicates an average over the IC and $ \Delta {\bf r}_i^2(t) = {\bf r}_i^2(t=t) - {\bf r}_i^2(t=0)$ is the absolute value of squared displacement of particle $i$ in such time interval). Since this quantity is calculated by averaging over different trajectories originating from a fixed inital configuration, it represents the tendency of the particles to move from such initial configuration and thus reflects the constraints imposed by the initial structure on dynamics. We used a short timescale $t=0.1\cdot t^*$ and also $t=t^*$ (the spatial distribution of propensity does not depend on such choice since it reflects a property of the initial structural configuration). $t^*$ represents a characteristic timescale of this system: The timescale of maximum inhomogeneous (non-gaussian) behavior and marks the end of the caging regime and the onset of the $\alpha$-relaxation \cite{donati_98}. $t^*$ depends strongly on temperature (it increases quickly towards $T_{\rm c}$ \cite{donati_98,appignanesi_05}) and has been shown to provide an estimate of the MB residence time \cite{donati_98,appignanesi_05}.
For each $T$ we stored 160 configurations separated one another by $10\cdot t^*$. The ICs were performed from these stored configurations and each one comprised 200 trajectories. We then averaged results from all the 160 ICs (this averaging is neccessary since we have already demonstrated that a given IC represents the structural constraints of the local MB where the systems is temporarily confined, and there is a whole distribution of MB sizes and MB residence times \cite{PRL_2}).   
 
In turn, we also calculated the distribution of particle squared displacements (without any previous IC averaging):
%$ \Delta {\bf r}_i^2(t) = {\bf r}_i^2(t=t) - {\bf r}_i^2(t=0)$
$ \Delta {\bf r}_i^2(t) = {\bf r}_i^2(t=t) - {\bf r}_i^2(t=0)$
% $4 \pi r^2 G_{\rm s}(r,t) = N^{-1} \sum_{i=1}^N \langle \delta(r-|{\bf r}_i(t)-{\bf r}_i(0)|) \rangle$,
%$4 \pi r^2 G_{\rm{s}}^2(r,t) = \frac{1}{N} \cdot \frac{1}{{\rm d}r} \sum_{i=1}^{N} \langle \int^{r2+\frac{{\rm d}r}{2}}_{r-\frac{{\rm d}r}{2}} \delta \left( r^2 - ( {\bf r}_i(t) - {\bf r}_i(0) ^2 \right) {\rm d}r \rangle$
%where $\delta$ is the Dirac delta function and ${\bf r^2}(t)$ is square of the value of the position of particle $i$ at time $t$ 
(we note that if instead of employing the squared displacements we used the linear displacements, the corresponding distribution would be the self part of the van Hove correlation function: $4 \pi r^2 G_{\rm{s}}(r,t) = \frac{1}{N} \cdot \frac{1}{{\rm d}r} \sum_{i=1}^{N} \langle \int^{r+\frac{{\rm d}r}{2}}_{r-\frac{{\rm d}r}{2}} \delta \left( r - | {\bf r}_i(t) - {\bf r}_i(0) | \right) {\rm d}r \rangle$
where $\delta$ is the Dirac delta function). At any given temperature, this function was calculated by considering explicitly the displacement of each particle within each of the 200 trajectories and for each of the 160 ICs. Given the large statistics provided by using many ICs, the resulting squared-displacement distribution (SDD) is identical to the result when the function is calculated in the usual way, that is, by collecting displacements from many independent single trajectories (no IC method at all), as is done to get van Hove function. 
We note that the SDD and the propensity distribution (PD) we calculate share the same data (and hence, have the same mean value). However, the SDD calculation does not imply any averaging (each particle from each trajectory contributes individually to the SDD, which amounts to $800 \times 200 \times 160$ total data: 800 A particles, 200 IC trajectories and 160 IC ensembles). On the contrary, the PD implies an isoconfigurational averaging (that is, the contribution of each particle is previously averaged over the 200 trajectories of the corresponding IC to get a propensity value and thus, the total number of data of the PD is $800 \times 160$).

\section{Results}

\begin{figure}[tb]
\includegraphics[width=0.9\linewidth]{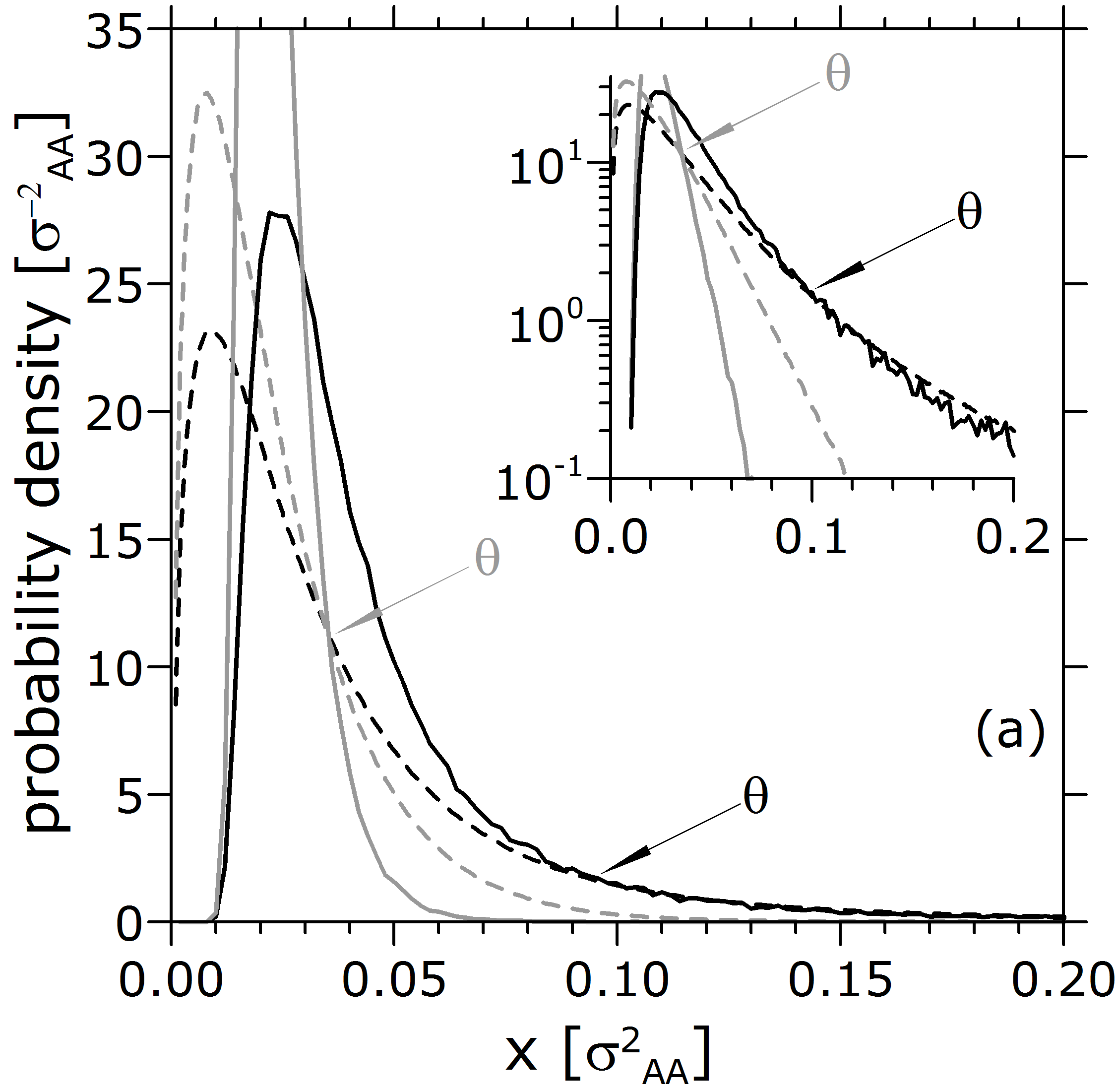}
\includegraphics[width=0.9\linewidth]{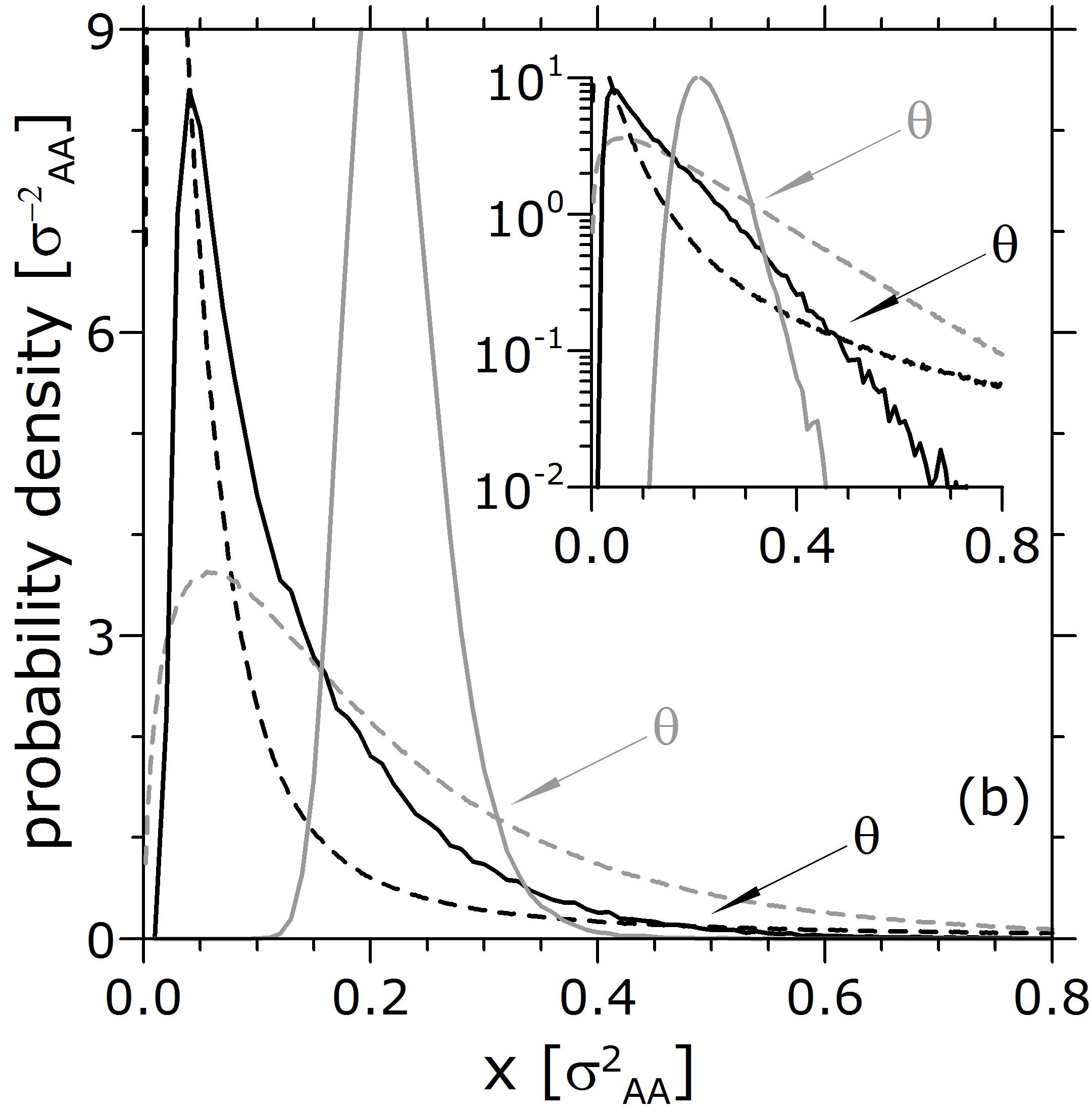}
\caption{
Comparison of the SDD (squared-displacement distribution, dashed lines) and the PD (propensity distribution, solid lines) for $T=2.0$ (gray) and $T=0.446$ (black). a) $t=0.1 \cdot t^*$, b) $t=t^*$. $\theta$ is the second intersection between the SDD and the PD. The abscisas $X$ is squared displacement or propensity of a particle A. The tendency of the propensity distribution to follow the long tail decay of the SDD at low temperatures is remarkable. The insets present the same curves in logarithmic representation.
}
\label{fig1}
\end{figure}

\begin{figure}[tb]
\includegraphics[width=0.9\linewidth]{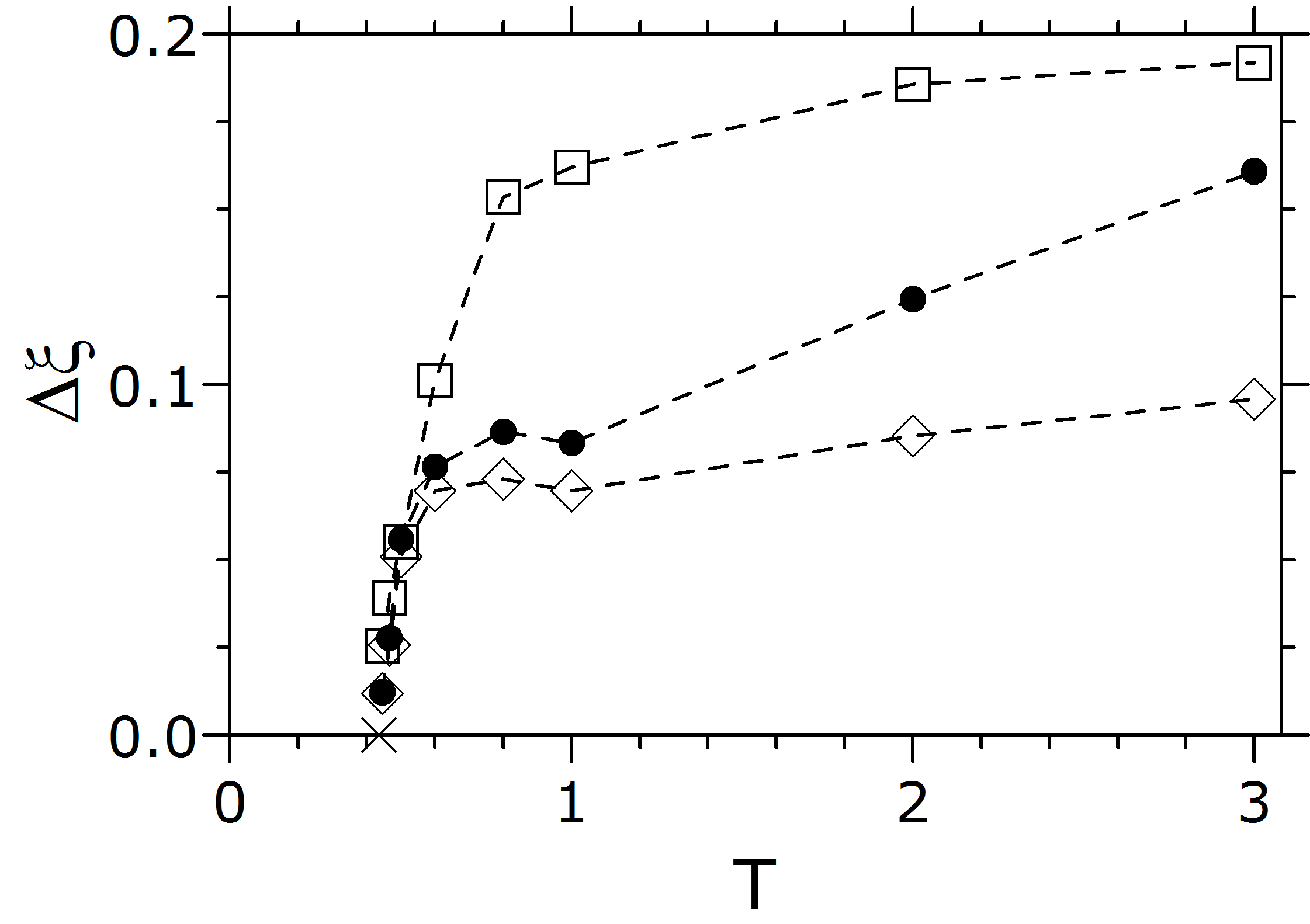}
\caption{
Temperature dependence of the excess $\Delta \xi$ of the squared-displacement distributions (SDD) over the propensity distributions (PD) after $\theta$. Circles: $t=0.1 \cdot t^*$, squares: $t=t^*$. The cross represents the mode-coupling temperature $T_{\rm c}$. The open diamonds represent the same case of  $t=0.1 \cdot t^*$ (circles) but when $\theta$ is the value above which the SDD curve, at each $T$, presents an area equal to $0.1$.
}
\label{fig2}
\end{figure}

In Figure \ref{fig1} we show the comparison of the PD and the SDD for two selected temperatures, $T=2.0$ and $T=0.446$, and for $t=0.1 \cdot t^*$ and $t=t^*$.
From such figure it is evident that at high temperatures the SDD and the PD are very different, with the latter being sharp and decaying quickly while the former is broader, presenting a large tail at large displacements (it is notable that a simple IC averaging process produces such a great change in the shape of the distribution function). However, we can also learn that at low $T$ the PD is much broader (as already noted \cite{widmer-cooper_04}) and, more importantly, that the PD tends to closely follow the decay of the SDD at large $\Delta {\bf r}^2$-values. It is obvious that the PD should tend to decay earlier than the corresponding SDD since it is calculated on IC-averaged values for the particles. A high propensity particle would present a whole distribution of displacements for the 200 IC trajectories (in some trajectories it would exhibit very large displacements while in others the mobility would be more modest and thus, it would not constitute one of the most mobile particles of the trajectory). The largest of these particle displacements are the ones that would contribute to the tail of the SDD. However, such high propensity particle would contribute to the PD with a lower value as provided by the averaging over all the trajectories of the IC. Thus, the conspicuous tendency of the PD to resemble the long $\Delta {\bf r}^2$ decay of the SDD at low $T$ speaks of the fact that the sets of mobile particles tend to be similar for all the trajectories of the IC. In contrast, the sharp distribution at high $T$ indicates that the sets of mobile particles differ significantly from one IC trajectory to another.
We thus decided to calculate the $\Delta{\bf r}^2$ value of the second intersection between the SDD and the PD (the intersection at long $\Delta{\bf r}^2$-values), which we call $\theta$ (see Figure \ref{fig1} for an estimation of this value for a couple of extreme temperatures). %The inset of Figure \ref{fig1} displays the temperature dependence of ${\bf r}_{\rm long}$. The rather temperature dependent behavior yielded could be due to the fact that $t^*$ represents a characteristic time. 
In turn, this value (that signals very mobile particles) allows one to calculate the ``excess'' of the SDD over the PD, which we define as $\Delta \xi = \int_\theta^{\infty} {\rm SDD} \cdot {\rm d}\Delta{\bf r}^2$ - $\int_\theta^{\infty} {\rm PD} \cdot {\rm d}P$ and whose $T$ dependence we display in Figure \ref{fig2}.
 
The value of $\Delta \xi$ measures how different from each other are the sets of mobile particles exhibited by the different trajectories of the IC. Ideally, if the long tail decay of the PD matched perfectly the SDD function, this would mean that the more mobile particles would be the same and would move equally in all the trajectories of the IC. From Figure \ref{fig2} we can learn that at high temperatures the trajectories of the IC present different sets of mobile particles, a fact that speaks of the existence of plenty of different relaxation pathways. However, below $T \approx 1$ this fuction begins to decay quickly and somehow tends to collapse at low temperature values approaching $T_{\rm C}$. This means that the sets of mobile particles for each IC trajectory tend to be more and more similar to each other as temperature lowers. In other words, as $T$ decreases, the trajectory needs to reach a simmilar relaxation event. From our previous results linking propensity with d-clusters and MBs \cite{PRL_2}, we know that the different trajectories of the IC (diverging trajectories from a common structural origin) explore the local MB in different ways in order to find a d-cluster to perform a transition to another MB, an event that triggers the $\alpha$ relaxation and decorrelates the PD pattern. In the light of the present results it is evident that, as $T$ decreases the trajectory has a fewer number of relaxation pathways available, that is, it can reach fewer events capable of producing the exit from the local MB. It is interesting that the change in behavior at $T \approx 1$ evident from Figure \ref{fig2} coincides with the change in exploration of the PEL from a free-diffusion to a ``landscape influenced'' regime, as found previously \cite{sastry_98}. In turn, the abrupt decay as $T$ is furthermore decreased is also consistent with the estimated onset of a ``landscape dominated'' regime around $T_{\rm c}=0.435$ within such description \cite{sastry_98}. To test the robustness of our finding, we also performed a calculation for a larger system of $N=8000$ particles and obtained very similar results. We also used a different definition of the value $\theta$ as the value beyond which, at each $T$, the area of the SDD amounts to $0.1$ (that is, the $10 \%$ of the more mobile particles). For each $T$ and for the IC of length $t=0.1 \cdot t^*$ we calculated the excess area of the SDD curve over the corresponding PD curve. The results, also displayed in Figure \ref{fig2} show the same tendency as the other two curves. Other choices of $\theta$ also gave similar results.

\begin{figure}[tb]
\includegraphics[width=0.9\linewidth]{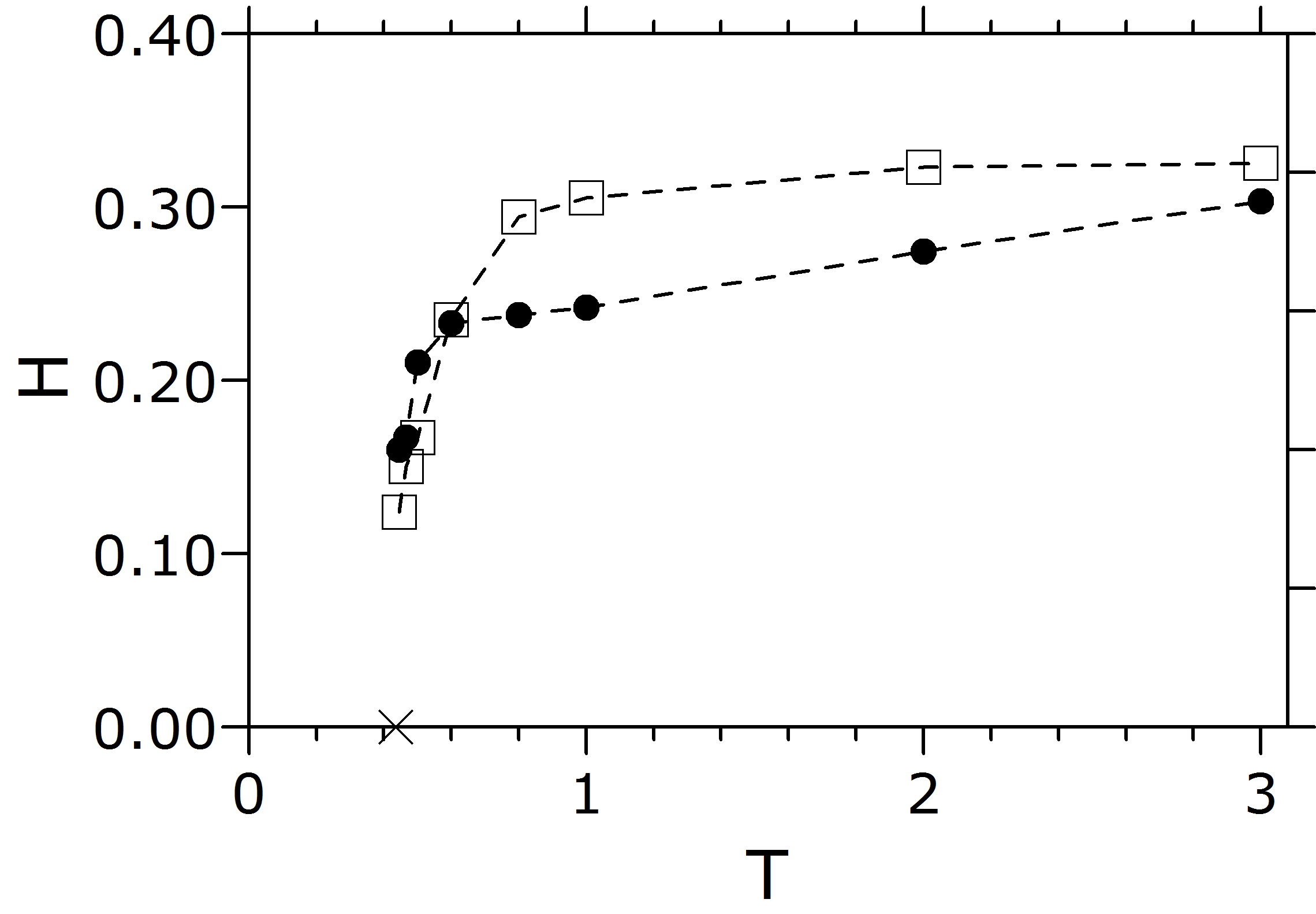}
\caption{
Shannon entropy $H$ as a function of temperature. Circles: $t=0.1 \cdot t^*$, squares: $t=t^*$.
% for the case of Figure \ref{fig1} and Figure \ref{fig2}.
The cross represents the mode-coupling temperature $T_{\rm c}$.
}
\label{fig3}
\end{figure}

The confinement of the relaxation pathway suggested by Figure \ref{fig2} represents an ``entropic funnel'' for the dynamical trajectories which can also be made explicit with the calculation of the Shannon entropy, $H=\sum_{i} {p(i)\cdot \log{p(i)}}$, where $p(i)$ represents the probability that a given particle $i$ has exceeded the mobility value $\theta$ in the different trajectories of the IC (calculated as the number of trajectories when such threshold displacement has been exceeded by the particle divided by the number of trajectories of the IC). The function $p(i)\cdot \log{p(i)}$ presents a single maximum at probability $p=0.5$ and is null at $p=0.0$ and $p=1.0$. Thus a high value of Shannon entropy $H$ would imply that most particles would present intermediate values in their probability of being mobile (of exhibiting very large displacements, larger than $\theta$). In turn, a low value of $H$ would mean that the particles would present high or low probabilities for such large mobility, without presence of significant intermediate behaviors. In other words, a low $H$ value would mean that the sets of mobile particles (the ones comprised by the particles which exceed the value $\theta$) would be similar for all the trajectories of the IC. Figure \ref{fig3} shows that, consistently with the results of Figure \ref{fig2}, as $T$ is lowered in the supercooled regime, the Shannon entropy begins to change below $T \approx 1$ and decays quickly as we approach $T_{\rm c}$.

\section{Conclusions}
By means of extensive computations and calculations of detailed and specific dynamical quantities, our results shed a new additional light on the dynamical slowing down that occurs within the supercooled regime in a glass-forming system, while reconciling dynamical real-space, landscape-based and structural-propensity descriptions. We have shown, by computing the excess of particle-mobility distributions over the corresponding propensity distributions, that while the dynamic trajectory has a rich menu of available relaxation alternatives at high temperatures (many different relaxing pathways made up by different mobile particles), glassy dynamics emerges at low temperatures since the relaxation pathway becomes increasingly confined to a common set of mobile particles. In other words, at high temperature there are many mobile particles (that can promote many different relaxation events) while the relaxation events can only choose from a reduced set of possible mobile particles as temperature decreases. This ``entropic funnel'' of relaxation pathways is furthermore made evident with the abrupt decay of the associated Shannon entropy.

%\end{document}

\section{Acknowledgments}
We have benefited from enlightening conversations with Prof. Pablo Debenedetti. Financial support from MinCyT and CONICET is gratefully acknowledged.

\end{document}